\documentclass[twocolumn,showpacs,prl]{revtex4}
\usepackage{graphicx,color,amsmath,amssymb}

\begin{document}
\title{Confinement and Superfluidity in one-dimensional
Degenerate Fermionic Cold Atoms}
\author{P. Lecheminant}
\affiliation{Laboratoire de Physique Th\'eorique et
Mod\'elisation, CNRS UMR 8089,
Universit\'e de Cergy-Pontoise, Site de Saint-Martin,
2 avenue Adolphe Chauvin,
95302 Cergy-Pontoise Cedex, France}
\author{E. Boulat}
\affiliation{Service de Physique Th\'eorique,
CEA Saclay, 91191 Gif Sur Yvette, France}
\author{P. Azaria}
\affiliation{
Laboratoire de Physique Th\'eorique des Liquides,
Universit\'e Pierre et Marie Curie, 4 Place Jussieu,
75256 Paris Cedex 05, France}
%%%%%%%%%%%%%%%%%%%%%%%%%%%%%%%%%%%%%%%%%%%%%%%%%%%%%%%%%%%%%%%%%%%%%%%%
\begin{abstract}
The physical properties of arbitrary half-integer spins $F = N - 1/2$
fermionic cold atoms trapped in a one-dimensional optical lattice are
investigated by means of a low-energy approach.  Two different
superfluid phases are found for $F \ge 3/2$ depending on whether a discrete
symmetry is spontaneously broken or not: an unconfined BCS pairing
phase and a confined molecular superfluid instability made of $2N$
fermions. We propose an experimental distinction between these phases
for a gas trapped in an annular geometry.
The confined-unconfined transition is shown to belong to the
${\mathbb{Z}}_N$ generalized Ising universality class.  We discuss on
the possible Mott phases at $1/2N$ filling.
\end{abstract}
%%%%%%%%%%%%%%%%%%%%%%%%%%%%%%%%%%%%%%%%%%%%%%%%%%%%%%%%%%%%%%%%%%%%%%%
\maketitle

In the recent past, spectacular experimental progress has allowed to
cool down alkali atoms below Fermi temperatures offering the promising
perspective to study strongly correlated electronic effects, such as
high-temperature superconductivity, in a new context
\cite{hofstetter}.  On top of strong correlations, ultracold atomic
systems provide also an opportunity to investigate the effect of spin
degeneracy since in general alkali fermionic atoms have (hyperfine)
spins $F> 1/2$ in their lowest hyperfine manifold.  Simplest examples
are spin-3/2 atoms such as $^{9}$Be, $^{132}$Cs, and $^{135}$Ba atoms.
In this Letter, we elucidate the generic physical features of
half-integer spins $F= N - 1/2$ fermionic cold atoms in the particular
case of a one-dimensional optical lattice.  The low-energy physical
properties of these systems are known to be described by a
Hubbard-like Hamiltonian \cite{ho}:
\begin{equation}
{\cal H} = -t \sum_{i,\alpha} \left[c^{\dagger}_{\alpha,i} c_{\alpha,
i+1} + {\rm H.c.} \right]+ \sum_{i,J} U_J \sum_{M=-J}^{J}
P_{JM,i}^{\dagger} P_{JM,i},
\label{hubbardSgen}
\end{equation}
where $c^{\dagger}_{\alpha,i}$ ($\alpha = 1,..., 2N$) is the fermion
creation operator corresponding to the $2F+1=2N$ atomic states.  The
pairing operators in Eq. (\ref{hubbardSgen}) are defined through the
Clebsch-Gordan coefficient for two indistinguishable particles:
$P^{\dagger}_{JM,i} = \sum_{\alpha \beta} \langle JM|F,F;\alpha,
\beta\rangle c^{\dagger}_{\alpha,i}c^{\dagger}_{\beta,i}$.  The
interactions are SU(2) spin-conserving and depend on $U_J$ parameters
corresponding to the total spin of two spin-$F$ particles which takes
only even integers value due to Pauli's principle: $J=0,2,...,2N-2$.
Even in this simple scheme the interaction pattern is still involved.
It is thus highly desirable to focus on a much simpler paradigmatic
model that incorporates the relevant physics of higher-spin
degeneracy.  In this respect, we consider a two coupling-constant
version of model (\ref{hubbardSgen}) with $U_2 = ... = U_{2N-2} \ne
U_0$:
\begin{equation}
{\cal H} = -t \sum_{i,\alpha} [c^{\dagger}_{\alpha,i} c_{\alpha,i+1} +
{\rm H.c.} ]+ \frac{U}{2} \sum_i \rho_i^2 + V \sum_i \pi^{\dagger}_i
\pi_i,
\label{hubbardS}
\end{equation}
$\rho_i = \sum_{\alpha} c^{\dagger}_{\alpha,i} c_{\alpha,i}$ being the
particle number operator and $U= 2 U_2$, $V = U_0 - U_2$.  In
Eq. (\ref{hubbardS}), the singlet BCS pairing operator for spin-$F$
fermions is $\pi^{\dagger}_i = \sqrt{2N} P^{\dagger}_{00,i} =
c^{\dagger}_{\alpha,i} {\cal J}_{\alpha \beta} c^{\dagger}_{\beta,i}$,
the matrix ${\cal J}$ being the natural generalization of the familiar
antisymmetric tensor $\epsilon = i \sigma_2$ to spin $F > 1/2$.  Such
a singlet-pairing operator has been extensively studied in the context
of two-dimensional frustrated quantum magnets \cite{sachdev}.  When
$V=0$, i.e. $U_0 = U_2$, model (\ref{hubbardS}) corresponds to the
SU($2N$) Hubbard model.  The Hamiltonian (\ref{hubbardS}) for $V \ne
0$ still displays a large symmetry since it is invariant under the
Sp($2N$) group which consists of unitary matrices $U$ that satisfy
$U^* {\cal J} U^{\dagger} = {\cal J}$ \cite{wusp}.  In the spin
$F=1/2$ case, model (\ref{hubbardS}) reduces to the SU(2) Hubbard
chain since SU(2) $\simeq$ Sp(2).  In the spin-3/2 case, models
(\ref{hubbardS}) and (\ref{hubbardSgen}) have an exact SO(5) $\simeq$
Sp(4) symmetry \cite{wusp}.  This Sp($2N$) symmetry considerably
simplifies the problem but may appear rather artificial.  However, we
expect that, for generic and small interactions, the original SU(2)
spin-rotational invariance will be dynamically enlarged at
sufficiently low energy \cite{balents,boulat}.  A second reason to
consider the Sp($2N$) symmetric model (\ref{hubbardS}) stems from the
fact that the Sp($2N$) and SU(2) groups share the same center, the
${\mathbb{Z}}_2$ group. Moreover, the striking physical properties of
the system rely on the existence of a ${\mathbb{Z}}_N$ symmetry which
is {\it also} a symmetry of the SU(2) model (\ref{hubbardSgen}).  This
${\mathbb{Z}}_N$ symmetry, that simply amounts to a global redefinition 
of the fermion phase, is properly defined as the coset
between the center ${\mathbb{Z}}_{2N}$ of the SU($2N$) group and the
center ${\mathbb{Z}}_2$ of the Sp($2N$) or SU(2) one:
${\mathbb{Z}}_{N}$ = ${\mathbb{Z}}_{2N}$/${\mathbb{Z}}_2$ with
\begin{equation}
{\mathbb{Z}}_{2N} : c_{\alpha,i} \rightarrow \mbox{e}^{in\pi/N}
c_{\alpha,i} \; , n=0, ....,2N-1,
\label{Z2Nsymmetry}
\end{equation}
the ${\mathbb{Z}}_2$ symmetry being $c_{\alpha,i} \rightarrow -
c_{\alpha,i}$.  The ${\mathbb{Z}}_N$ symmetry, defined by
Eq. (\ref{Z2Nsymmetry}) with $n=0,...,N-1$, provides an important
physical ingredient not present in the $F=1/2$ case.
%While any order parameter has to be invariant under Z2 $c\rightarrow -c$,
%it can nevertheless carry a non zero ZN charge.
The stabilization of a quasi-long-range BCS phase for $F>1/2$ requires
the spontaneous breaking of this ${\mathbb{Z}}_N$ symmetry since the
singlet pairing $\pi_i^{\dagger}$ is \emph{not} invariant under this
symmetry.  Since ${\mathbb{Z}}_N$ may be also viewed as a discrete
subgroup of the global U(1) charge symmetry (see
Eq. (\ref{Z2Nsymmetry})), it is tempting to interpret, for $F>1/2$,
the breaking of ${\mathbb{Z}}_N$ as the reminiscence of the
spontaneous global U(1) charge breaking that characterizes the BCS
phase in higher dimensions.  In contrast, if ${\mathbb{Z}}_N$ is not
broken, the BCS instability is suppressed and the
leading superfluid instability, which has to be a singlet 
under the ${\mathbb{Z}}_N$ symmetry,  is a molecular object
made of $2N$ fermions.
In the following, the delicate competition between these
superfluid instabilities will be investigated by means of a low-energy
approach.

-{\it Phase diagram.}  The low-energy effective field theory
associated with Eq. (\ref{hubbardS}) is obtained, as usual, from the
continuum description of the lattice electronic operators in terms of
right and left moving Dirac fermions: $c_{\alpha,i}/\sqrt{a_0}
\rightarrow R_{\alpha}(x) \mbox{e}^{ik_F x} + L_{\alpha}(x)
\mbox{e}^{-ik_F x}, x= i a_0,$ $a_0$ being the lattice spacing and
$k_F$ is the Fermi momentum \cite{bookboso}.  Away from half-filling
(i.e. $N$ atoms per site), it separates into two commuting density and
spin pieces: ${\cal H} = {\cal H}_d + {\cal H}_s$ with $[{\cal H}_d,
  {\cal H}_s] = 0$.  The U(1) density sector is described by a bosonic
field $\Phi$ and its dual $\Theta$ whose dynamics is governed by the
free-boson Hamiltonian:
\begin{equation}
{\cal H}_d = \frac{v}{2} \left[\frac{1}{K_d} \left(\partial_x \Phi
\right)^2 + K_d \left(\partial_x \Theta \right)^2 \right],
\label{lutt}
\end{equation}
where $v = v_{\textsc{f}} \left[1 + (2 V + U N(2N -1))/(N\pi
v_{\textsc{f}})\right]^{1/2}$ ($v_{\textsc{f}} = 2t a_0 \sin
(k_{\textsc{f}} a_0)$ being the Fermi velocity) and $K_d = \left[1 +
(2 V + U N(2N -1))/(N\pi v_{\textsc{f}})\right]^{-1/2}$ are the
Luttinger parameters.  The conserved quantities in this U(1) sector
are the total particle number and current: ${\cal N} = \int d x \;
(R^{\dagger}_{\alpha} R_{\alpha} + L^{\dagger}_{\alpha} L_{\alpha}) =
\sqrt{2N/\pi} \int d x \; \partial_x \Phi$ and ${\cal J}= \int d x \;
(- R^{\dagger}_{\alpha} R_{\alpha} + L^{\dagger}_{\alpha} L_{\alpha})=
\sqrt{2N/\pi} \int d x \; \partial_x \Theta$ respectively.  For
incommensurate fillings, the density degrees of freedom are massless
and display metallic properties in the Luttinger liquid universality
class \cite{bookboso}.  All non-trivial physics is encoded in the spin
part of the Hamiltonian:
\begin{eqnarray}
{\cal H}_s &=& \frac{2\pi v_s}{2N + 1} \left[ I^a_{\parallel R}
I^a_{\parallel R} + I^i_{\perp R} I^i_{\perp R} + R \rightarrow L
\right] \nonumber \\ &+& g_{\parallel} I^a_{\parallel R}
I^a_{\parallel L} + g_{\perp} I^i_{\perp R} I^i_{\perp L},
\label{spinham}
\end{eqnarray}
where $g_{\parallel} = - 2 (2V + N U)/N$, $g_{\perp} = 2 (2V - N U)/N$
and we have neglected a velocity anisotropy.  The Hamiltonian
(\ref{spinham}) describes a SU($2N$)$_1$ conformal field theory (CFT)
perturbed by a marginal current-current interaction.  In
Eq. (\ref{spinham}), the currents $I^A_{R(L)}$, $A=(1,...,4N^2-1)$, of
the SU($2N$)$_1$ CFT have been decomposed into $\parallel$ and $\perp$
parts $I^A= (I^a_{\parallel}, I^i_{\perp})$ with respect to the
Sp($2N$) symmetry of the lattice model (\ref{hubbardS}).  The currents
$I^a_{\parallel R(L)}, a=1,..,N(2N+1)$ generate the Sp($2N$)$_1$ CFT
symmetry and can be simply expressed in terms of the chiral Dirac
fermions: $I^a_{\parallel R} = R_{\alpha}^{\dagger} T^a_{\alpha \beta}
R_{\beta}$, $T^a$ being the generators of Sp($2N$) in the fundamental
representation.  The remaining SU($2N$)$_1$ currents are written as
$I^i_{\perp R} = R_{\alpha}^{\dagger} T^i_{\alpha \beta}
R_{\beta},i=1,..,2N^2 - N - 1$ (similar expressions hold for the left
currents).  The next step of the approach is to consider a description
which singles out the ${\mathbb{Z}}_N$ symmetry of the lattice model
(\ref{hubbardS}) discussed above.  To this end, we shall use the
quantum equivalence \cite{altschuler}: U(2N)$_1$ $\rightarrow$ U(1)
$\times$ Sp($2N$)$_1$ $\times$ ${\mathbb{Z}}_N$, ${\mathbb{Z}}_N$
being the parafermion CFT which describes self-dual critical points of
two-dimensional ${\mathbb{Z}}_N$ Ising models \cite{para}.  This
conformal embedding provides us with a {\it non-perturbative} basis to
express any physical operator in terms of its density and spin degrees
of freedom which are described respectively by the U(1) and
Sp($2N$)$_1$ $\times$ ${\mathbb{Z}}_N$ CFTs.  The lattice
${\mathbb{Z}}_N$ symmetry is then captured, within this low-energy
approach, by an effective 2D ${\mathbb{Z}}_N$ model which is a
generalization to $N>2$ of the standard Ising model.  As in the $N=2$
case, these ${\mathbb{Z}}_N$ Ising models exhibit two phases described
by order and disorder parameters $\sigma_k$ and $\mu_k$, $k=1,..,N-1$,
which are dual to each other by means of the Kramers-Wannier (KW)
duality symmetry.  This duality transformation maps the
${\mathbb{Z}}_N$ symmetry, broken in the low-temperature phase
($\langle \sigma_k \rangle \ne 0$ and $\langle \mu_k \rangle = 0$),
onto a ${\tilde {\mathbb{Z}}}_N$ symmetry which is broken in the
high-temperature phase where $\langle \mu_k \rangle \ne 0$ and
$\langle \sigma_k \rangle = 0$.  At the critical point, the theory is
self-dual with a ${\mathbb{Z}}_N$ $\times$ ${\tilde {\mathbb{Z}}}_N$
symmetry and its universal properties are captured by the
${\mathbb{Z}}_N$ parafermion CFT \cite{para}.
\begin{figure}
\begin{center}
\includegraphics[width=0.8\linewidth]{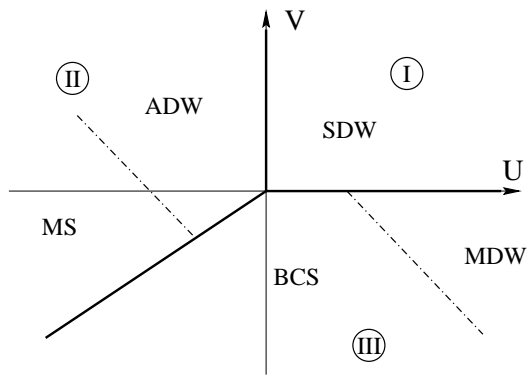}
\end{center}
\caption{ Phase diagram of model (\ref{hubbardS}).  Dashed lines
denote cross-over lines whereas the solid; this results in strong constraints on the possible
phases line marks the phase
transition in the ${\mathbb{Z}}_N$ universality class between phases
II and III.}
\vspace{-0.3cm}
\end{figure}
In the simplest $N=2$ case, there is a simple free-field
representation of the unperturbed SU(4)$_1$ CFT in terms of real
fermions which has been extensively used in the context of two-leg
ladders \cite{bookboso}. Introducing real fermions $\xi^{0}_{R,L}$ and
$\xi^{a}_{R,L} ,a=1,...,5$ to describe respectively the
${\mathbb{Z}}_2$ and SO(5)$_1$ $\simeq$ Sp(4)$_1$ CFTs, the
interacting part of Eq. (\ref{spinham}) becomes ${\cal H}_{{\rm
s}}^{{\rm int}} = g_{\parallel} (\xi^{a}_R \xi^{a}_L)^2 + g_{\perp}
\xi^{0}_{R} \xi^{0}_{L} \xi^{a}_R \xi^{a}_L$.  The latter model has
been studied recently to describe a SO(5) symmetric two-leg ladder
\cite{controzzi}.  For generic $N$, the phase diagram of
Eq. (\ref{spinham}) can be elucidated by means of a two-loop
renormalization group (RG) analysis. As depicted in Fig. 1 it consists
of three regions \cite{comment}.  Region I is a generalized
spin-density-wave (SDW) phase which is obtained when $U$ and $V$ are
positive.  In that case, $g_{\perp,\parallel} \rightarrow 0$ in the
infrared (IR) limit and the interaction is marginal irrelevant.  Up to
a spin-velocity anisotropy, the low-energy properties of this phase
are similar to that of the repulsive SU($2N$) Hubbard chain with $2N
-1$ gapless spin excitations \cite{affleck,assaraf}.  In contrast, a
spin gap opens in the two remaining regions which are distinguished by
the ${\mathbb{Z}}_N$ symmetry.  In phase II, defined by $U<0$ and $V >
N U/2$, the RG flow in the far IR limit is attracted along a special
symmetric ray $g_{\parallel} = g_{\perp} = g^{*} > 0$ where the
interacting part of the Hamiltonian (\ref{spinham}) can be rewritten
in a SU($2N$) invariant form:
\begin{equation}
{\cal H}_{{\rm s}}^{{\rm int}} = g^{*}\left( I^a_{\parallel R}
I^a_{\parallel L} + I^i_{\perp R} I^i_{\perp L} \right) = g^{*}
I^A_{R} I^A_{L}.
\label{su2NGN}
\end{equation}
The Hamiltonian (\ref{su2NGN}), which governs the IR properties of
phase II, takes the form of the SU($2N$) Gross-Neveu (GN) model which
is an integrable massive field theory \cite{andrei}.  It is
instructive to have a simpler understanding of the spin-gap formation
from the underlying ${\mathbb{Z}}_N$ Ising model.  The
${\mathbb{Z}}_N$ and ${\tilde {\mathbb{Z}}}_{N}$ symmetries that
define the low-T and high-T phases of this Ising model admit a
representation in terms of the fermions $(R,L)_{\alpha}$.  Indeed, the
${\mathbb{Z}}_{N}$ group (\ref{Z2Nsymmetry}) can be generated in this
chiral fermion basis with help of the unitary operator ${\cal U} =
\mbox{e}^{i \pi {\cal N}/N}$: ${\cal U} R(L)_{\alpha} {\cal
U}^{\dagger} = \mbox{e}^{i \pi /N}R(L)_{\alpha}$.  Similarly, the dual
${\tilde {\mathbb{Z}}}_{N}$ symmetry can be defined by ${\tilde {\cal
U}} = \mbox{e}^{i \pi {\cal J}/N}$: ${\tilde {\cal U}}
R(L)_{\alpha}{\tilde {\cal U}}^{\dagger} = \mbox{e}^{\mp i \pi
/N}R(L)_{\alpha}$.  The ground state of model (\ref{su2NGN}) displays
long-range order associated with the order parameter ${\rm Tr}(g) =
R^{\dagger}_{\alpha}L_{\alpha} \mbox{e}^{i\sqrt{2\pi/N} \Phi}$.  In
phase II, we find that the ${\mathbb{Z}}_{2N}={\mathbb{Z}}_2 \times
{\mathbb{Z}}_{N}$ symmetry remains unbroken while ${\tilde
{\mathbb{Z}}}_{N}$ is spontaneously broken.  The ${\mathbb{Z}}_{N}$
Ising model thus belongs to its high-T phase and a spectral gap is
formed.  In the second spin-gapped phase (III) of Fig. 1, defined by
$V<0$ and $V < N U/2$, the RG flow is now attracted along the
asymptote: $g_{\parallel} = - g_{\perp} = g^{*} > 0$.  In that case,
the interacting part of the IR Hamiltonian becomes
\begin{equation}
{\cal H}_{{\rm s}}^{{\rm int}} = g^{*}\left( I^a_{\parallel R}
I^a_{\parallel L} - I^i_{\perp R} I^i_{\perp L} \right),
\label{su2NGNdual}
\end{equation}
which can be recast as a SU($2N$) GN model (\ref{su2NGN}) by means of
a $duality$ transformation ${\cal D}$ on the fermions: ${\cal
D}R(L){\cal D}^{-1} = {\tilde R(\tilde {L})}$ with ${\tilde
R}_{\alpha} = {\cal J}_{\alpha \beta} R^{\dagger}_{\beta}$ and $
{\tilde L}_{\alpha}= L_{\alpha}$.  This transformation acts on the
currents as: ${\tilde I}^a_{\parallel R(L)} = I^a_{\parallel R(L)}$
and $ {\tilde I}^i_{\perp R(L)}= - (+) I^i_{\perp R(L)}$ so that
${\cal D}$ indeed maps (\ref{su2NGNdual}) onto (\ref{su2NGN}).
Besides the opening of a spectral gap, we thus find that phase III
possesses a hidden symmetry at low energy i.e.  a $\widetilde{SU}(2N)$
symmetry generated by the dual currents $({\tilde I}^a_{\parallel
R(L)},{\tilde I}^i_{\perp R(L)})$.  In fact, one has ${\cal D} {\cal
U} {\cal D}^{-1}= {\tilde {\cal U}}$ so that ${\cal D}$ identifies to
the KW duality of the ${\mathbb{Z}}_N$ Ising model.  In phase III, the
latter model thus belongs to its low-T phase and the ${\mathbb{Z}}_N$
symmetry is spontaneously broken whereas ${\tilde {\mathbb{Z}}}_{2N} =
{\mathbb{Z}}_2 \times {\tilde {\mathbb{Z}}}_{N}$ remains unbroken.  In
summary, the existence of these two distinct spin-gapped phases is a
non-trivial consequence of higher-spin degeneracy and does not occur
in the $F=1/2$ case.  The emergence of the spin-gap stems from the
spontaneous breakdown of the ${\mathbb{Z}}_N$ or ${\tilde
{\mathbb{Z}}}_N$ discrete symmetries.  As we shall see now, these
symmetries are central to the striking physical properties displayed
by these phases.

-{\it Spin superfluidity}. The low-energy properties of the spin
sector of phase II can be extracted from the integrability of the
SU($2N$) GN model (\ref{su2NGN}).  Its spectrum consists into $2N -1$
branches that transform in the SU($2N$) representations \cite{andrei}.
These eigenstates are labelled by quantum numbers associated with the
conserved quantities of the SU($2N$) low-energy symmetry:
$Q_{\parallel}^a = \int dx \; (I^a_{\parallel L} + I^a_{\parallel
  R})$, $a= (1,..., N)$ and $Q_{\perp}^i = \int dx \; (I^i_{\perp L }
+ I^i_{\perp R })$, $i= (1,..., N-1)$.  Due to the Sp($2N$) symmetry
of model (\ref{hubbardS}), the $Q_{\parallel}^a$ numbers are conserved
whereas the $Q_{\perp}^i$ charges are only good quantum numbers at low
energy.  The spin spectrum in phase III can be obtained from the
duality symmetry ${\cal D}$ and consists into $2N -1$ branches which
transform in the representations of the dual group
$\widetilde{SU}(2N)$.  The dual quantum numbers are now given by:
${\tilde Q}_{\parallel}^a = Q_{\parallel}^a$ and ${\tilde Q}_{\perp}^i
= \int dx \; ({\tilde I}^i_{\perp L } + {\tilde I}^i_{\perp R }) =
\int dx \; (I^i_{\perp L } - I^i_{\perp R }) = J_{\perp}^i$.  We thus
observe that the low-lying excitations in phase III carry quantized
spin $currents$ in the ``$\perp$'' direction.  In this sense, the
phase III might be viewed as a partially spin-superfluid phase.

-{\it Confinement}. We shall now determine the nature of the dominant
electronic instabilities of the different phases of Fig. 1.  To this
end let us consider operators ${\cal O}_{n,j}$ which carry particle
number $n$ and current $j$: $[{\cal N} ({\cal J}) , {\cal O}_{n,j}] =
n(j){\cal O}_{n,j}$.  Using the ${\mathbb{Z}}_{2N}$ and ${\tilde
  {\mathbb{Z}}}_{2N}$ generators, ${\cal U}$ and ${\tilde {\cal U}}$,
we find that ${\cal O}_{n,j}$ carry ${\mathbb{Z}}_{2N}$ and ${\tilde
  {\mathbb{Z}}}_{2N}$ charges $n$ and $j$ respectively.  In phase II,
the full ${\mathbb{Z}}_{2N} =$ ${\mathbb{Z}}_2\times$ ${\mathbb{Z}}_N$
symmetry (\ref{Z2Nsymmetry}) is unbroken so that it costs a finite
energy gap to excite states that either break the ${\mathbb{Z}}_2$ or
${\mathbb{Z}}_N$ symmetries.  The dominant instabilities must thus be
neutral under the ${\mathbb{Z}}_{2N}$ symmetry and the resulting order
parameters ${\cal O}_{n,j}$ are characterized by $n = 0 \; {\rm mod}
\; 2N$ and $j = 0 \; {\rm mod} \; 2$.  In particular, there is no
quasi-long-range BCS order in phase II since the lattice singlet
pairing operator $\pi_i^{\dagger}$ carries a charge $n=2$ under the
${\mathbb{Z}}_{2N}$ symmetry (\ref{Z2Nsymmetry}).  The
${\mathbb{Z}}_{2N}$ symmetry thus confines the electronic charge to
multiple of $2Ne~$ i.e.  the leading superfluid instability in phase
II is a composite object made of $2N$ fermions.  In this respect, the
dominant order parameters in phase II are: $\rho_{2k_{\textsc{f}}}=
L^{\dagger}_{\alpha} R_{\alpha}$ and $\Pi^{2N \dagger }_0 =
\epsilon^{\alpha_1 .. \beta_N} R^{\dagger}_{\alpha_1}...
R^{\dagger}_{\alpha_N}L^{\dagger}_{\beta_1}...L^{\dagger}_{\beta_N} $
which are, respectively, the $2k_{\textsc{f}}$ component of the atomic
density $\rho$ and the uniform component of the lattice
SU($2N$)-singlet superconducting instability made of $2N$ fermions:
$\Pi^{2N \dagger }(i) = \epsilon^{\alpha_1 .. \alpha_{2N}}
c_{\alpha_1,i}^{\dagger}..c_{\alpha_{2N},i}^{\dagger}$.  These orders
are power-law fluctuating: $\langle
\rho_{2k_{\textsc{f}}}^{\dagger}(x) \rho_{2k_{\textsc{f}}}(0) \rangle
\sim x^{-K_d/N}$, and $\langle \Pi^{2N \dagger }_0(x) \Pi^{2N}_0(0)
\rangle \sim x^{-N/K_d}$.  For $K_d < N$, the leading instability is
$\rho_{2k_{\textsc{f}}}$ which gives rise to an atomic-density wave
(ADW) phase whereas for $K_d > N$ a SU($2N$) molecular-superfluid (MS)
phase is stabilized (see Fig. 1) with order parameter $\Pi^{2N \dagger
}_0$.  The properties of phase III are obtained from those of phase II
with help of the duality symmetry ${\cal D} : ( {\cal N}
\leftrightarrow {\cal J}, {\mathbb{Z}}_N \leftrightarrow {\tilde
  {\mathbb{Z}}}_N, K_d \leftrightarrow 1/K_d)$.  Low-energy
excitations in phase III carry now $n = 0 \; {\rm mod} \; 2$ and $j =
0 \; {\rm mod} \; 2N$ since the symmetry ${\tilde {\mathbb{Z}}}_{2N} =
{\mathbb{Z}}_2 \times {\tilde {\mathbb{Z}}}_{N}$ remains unbroken.  We
find now the confinement of atomic currents and the emergence of a
quasi-long-range BCS pairing phase.  Under the duality ${\cal D}$
symmetry, the ADW phase is mapped onto a BCS phase for $K_d > 1/N$
with order parameter $\pi^{ \dagger}_0 = R^{\dagger}_{\alpha} {\cal
  J}_{\alpha \beta} L^{\dagger}_{\beta}$ whereas the MS phase is
mapped onto a molecular density-wave (MDW) phase with order parameter
${\bar \rho}^{2N}_{2Nk_{\textsc{f}}} = \epsilon^{\alpha_1..\beta_N}
{\cal J}_{\alpha_1 \gamma_1}...  {\cal J}_{\alpha_N \gamma_N}\;
R_{\gamma_1}...
R_{\gamma_N}L^{\dagger}_{\beta_1}...L^{\dagger}_{\beta_N} $ which
emerges when $K_d < 1/N$.  The spontaneous breaking of the
${\mathbb{Z}}_N$ symmetry (\ref{Z2Nsymmetry}) thus accounts for the
emergence of the BCS superfluid phase and the spin-superfluidity
phenomenon discussed above.  The possible occurrence of two different
superfluid phases II and III may be probed experimentally. Consider
for example a gas trapped in an optical potential of length $L$, with
an {\it annular} geometry and moving with tangential velocity
$V$. This amounts to imposing a total particle current in the system
$J= 4NV/V_0$ where $V_0 = h/m L$. In the superfluid phase III since
the low energy excitations carry currents $j = 0 \; {\rm mod} \; 2N$
we expect the total energy E(V) to display degenerate minima for
quantized velocities : $V_n = n V_0/2$ irrespective of the value of
the spin $F$.  In contrast in the phase II where currents are
unconfined, we expect the degenerate minima of E(V) at $V_n = n
V_0/2N$.

-{\it The ${\mathbb{Z}}_N$ phase transition}.  The nature of the
quantum phase transition between the two spin-gapped phases II and III
can be determined through the duality symmetry ${\cal D}$.  On the
self-dual line $g_{\perp}=0$, i.e. $2V = NU$, there is a separation of
the Sp($2N$) and ${\mathbb{Z}}_N$ degrees of freedom.  Though the
Sp($2N$) sector remains gapfull when $U < 0$, the effective
${\mathbb{Z}}_N$ Ising model is at its self-dual critical point and
governs the phase transition.  The ${\mathbb{Z}}_N$ quantum
criticality for $N=2,3$ may be revealed by considering the ratio
${\cal R}_N(x) = (\langle \pi^{ \dagger}_0 (x) \pi_0 (0)
\rangle)^{N^2} /\langle \Pi^{2N \dagger }_0(x) \Pi^{2N}_0(0)\rangle$
which displays a power-law decay with a \emph{universal} exponent: ${\cal
  R}_N(x) \sim x^{-2 N(N - 1)/(N + 2)}$.  For larger $N$ the phase
transition is non-universal.  For $N \ge 4$, a strongly relevant
perturbation is indeed generated which takes the form of the second
thermal operator $\epsilon_2$ of the ${\mathbb{Z}}_N$ CFT with scaling
dimension $12/(N+2)$ \cite{para}.  The resulting model is integrable
and the transition is either of first-order or in the U(1)
universality class depending on $N$ and the sign of the coupling
constant of $\epsilon_2$ \cite{fateev}.  In the $N=4$ case, i.e. a
special case of the Ashkin-Teller model, the ${\mathbb{Z}}_4$
criticality can emerge with the introduction of the interaction
$\lambda \sum_i (P_{00,i}^{\dagger} P_{00,i})^2$ which can eliminate
the operator $\epsilon_2$ by a fine-tuning of $\lambda$.

-{\it Mott phases}. At the commensurate $1/2N$ filling, i.e. one atom
per site, an umklapp term $ \sim \cos\big( \sqrt{8\pi N} \Phi\big)$ is
generated in the density sector and becomes relevant when $K_d < 1/N$
leading to the opening of a density gap.  We further distinguish
between three different Mott phases \cite{wucoment}.  The first one
lies in the SDW region I and is qualitatively similar to the one
encountered in the pure SU($2N$) Hubbard chain \cite{assaraf}. The two
others have a spin gap and can be distinguished with respect to the
confinement properties of atomic currents.  In the region II we find
for large enough $V$ a 2k$_{\textsc{f}}$-ordered ADW and spin-Peierls
ordering with a $2N$ ground-state degeneracy. In the region III the
cross-over line between BCS and MDW phases identifies to the Mott
transition line. At filling $1/2N$, while the BCS phase remains, the
MDW regime locks into a $2Nk_{\textsc{f}}$ MDW and displays a
dimerized bond ordering for all $N$. This Mott phase is unusual since
there is no one-particle density long-range fluctuation due to the
confinement of atomic currents. Remarkably enough we find that this
MDW Mott phase is the only gapped phase directly connected to the BCS
superfluid phase.


\begin{thebibliography}{10}
\bibitem{wucoment} The Mott phases in the $N=2$ case at
quarter-filling have also been found independently by C. Wu, private
communications.
\end{thebibliography}
\end{document}